\documentstyle[draft,twoside,fleqn,espcrc2,psfig,epsf]{article}

\newcommand{\AmS}{{\protect\the\textfont2
  A\kern-.1667em\lower.5ex\hbox{M}\kern-.125emS}}

\title{Random Surfaces and Lattice Gravity}

\author{Mark Bowick\address{Physics Department, 
        Syracuse University, \\ 
        Syracuse NY 13244-1130, \\
        U.S.A. }%
        \thanks{Research supported by the Department of Energy U.S.A under
        Contract No. DE-FG02-85ER40237 and by Syracuse University.}}
        
\begin{document}

\begin{abstract}
In this talk I review some of the recent developments in the field of
random surfaces and the Dynamical Triangulation 
approach to simplicial quantum gravity.  
In two dimensions I focus on the $c=1$ barrier
and the fractal dimension of two-dimensional quantum gravity coupled to 
matter with emphasis on the comparison of analytic predictions
and numerical simulations.
Next is a survey of the current understanding in 3 and 4 dimensions.
This is followed by a discussion of some problems in the statistical 
mechanics of membranes.
Finally I conclude with a list of problems for the future.
 
\end{abstract}

\maketitle

\section{Outline}

There are many approaches to formulating a discrete theory of quantum
gravity. In this talk I will focus on the dynamical triangulations
(DT) formulation of simplicial gravity. This was first developed in
the context of string theory and two-dimensional quantum gravity
($QG_2$) where the discrete approach has proven extremely powerful and
actually preceded continuum treatments.  By now we have considerable
confidence in the validity of the assumptions underlying the DT model
in the case of coupling of matter with central charge c less than
one. This confidence stems from the agreement of discrete, continuum
and numerical results.

The situation for central charge c greater than one was clarified in
the last year by David \cite{Davi97}.  After the Introduction a
discussion of this work will be the first part of this review.  This
is followed by a discussion of recent numerical tests of David's
proposal \cite{ThPe97}.

I will then move on to the fascinating issue of the intrinsic fractal
geometry of $QG_2$ and $QG_2$ coupled to conformal matter.  In any
theory of gravity it is natural to ask what effect quantum
fluctuations have on the structure of spacetime.  The most basic
question we could ask of any spacetime with a given topology is:
``What is its Hausdorff dimension?'' Our current knowledge of the
answer to this question will be reviewed, with emphasis on a
comparison of analytic predictions with recent numerical simulations.

The success of the DT approach to simplicial gravity in two dimensions
has inspired several groups to tackle large scale simulations of the
DT discretization of Euclidean Einstein-Hilbert gravity in 3 and 4
dimensions. Here the situation is much cloudier, both theoretically
and numerically. The current status will be summarized in the third
part of my talk.

Any rich idea usually has many unsuspected spin-offs.  The field of
random surfaces and non-critical string theory is intimately connected
with the physics of membranes. Some recent numerical results on new
phases of anisotropic membranes will be discussed as an example of the
fascinating interdisciplinary nature of the subject of random
surfaces.

Finally I present a list of challenging future problems that I think
are essential for progress in the field.  I hope that some of them
will be solved by the time of Lattice 98.

\section{Introduction}
String theories may be viewed as 2-dimensional Euclidean quantum
field theories with particular actions and matter content. From this
viewpoint they may also be considered as 2d{-}statistical mechanical
models on lattices with dynamical geometry. Statistical mechanical
models on fixed lattices often possess special critical points where
they are scale invariant.  Correlation functions of generic matter
fields $\Phi(x)$ reflect this scale invariance by transforming very
simply under scale transformations: $\langle
\Phi\left(\vec{r}\right)\Phi(0) \rangle \sim r^{-2\Delta^o_\Phi}$,
where $\Delta^{o}_{\Phi}$ is the scaling dimension of $\Phi$, or
equivalently its anomalous dimension in field theory. Combining scale
invariance with locality leads to the much larger symmetry of
conformal invariance. In two dimensions conformal invariance and
unitarity restrict the possible values of critical exponents because
the unitarisable representations of the associated Virasoro algebra
form a discrete series parameterized by a single real number {---} the
central charge $c$. The central charge determines the effect of scalar
curvature on the free energy of the model \cite{Gin88}.

For $c<1$ the only allowed values are
\begin{equation} 
\label{eq:minseries}
c = 1 - \frac{6}{m(m+1)}\quad \mbox{with}\quad m = 2,3,4 \ldots 
\end{equation}
It is also understood, through the classic work of KPZ \cite{KPZ88}, 
how scaling dimensions of fields are modified when the lattice becomes 
dynamical i.e. when the model is coupled to two-dimensional gravity.
The dressed scaling dimensions $\Delta_\Phi$ are determined solely by
$\Delta^{0}_{\Phi}$ and the central charge $c$ via
\begin{equation}
\Delta_\Phi - \Delta^o_{\Phi} = \frac{\Delta_{\Phi}(1-\Delta_\Phi)}
{1-\gamma_s\left(0\right)},
\end{equation}
where the string susceptibility exponent $\gamma_s$ describes the
entropy of random surfaces of fixed area $A$ {\em viz}:

\begin{equation}
\label{eq:fixedaz}
Z(A)={\rm e}^{\Lambda_{c}A} A^{\gamma_{s}-3}
\end{equation}
and
\begin{equation}
\label{eq:gams}
\gamma_s(h) = 2 - \frac{1-h}{12}\Big\{25-c+\sqrt{(1-c)(25-c)}\,\Big\}
\end{equation}
for a Riemann surface of genus $h$.
Eq.(\ref{eq:fixedaz}) implies the singularity structure
\begin{equation}
\label{eq:fixedcz}
Z(\Lambda)=\int^{\infty}_0 dA e^{-\Lambda A} Z(A) \sim 
(\Lambda_c-\Lambda)^{2-\gamma_s}
\end{equation}
as $\Lambda \rightarrow \Lambda_c$.
The renormalized continuum theory is obtained by tuning the 
cosmological constant $\Lambda$ to the critical cosmological constant 
$\Lambda_c$. In this limit the mean area $\langle A \rangle$ diverges like
$1/(\Lambda-\Lambda_c)$.
The linearity of $\gamma_s$ with genus $h$ given by Eq.(\ref{eq:gams})
has the remarkable  consequence that the partition function including the sum
over genus
\begin{equation}
\label{eq:topsum}
Z(\Lambda,\mu) = \sum^{\infty}_0 e^{-\mu h} \int^{\infty}_0 dA\; e^{-\Lambda A}
Z(A)
\end{equation} 
is actually a function not of two variables $\Lambda$ and $\mu$ but only
of the single scaling combination 
\begin{equation}
\label{eq:scalvar}
x=(\Lambda-\Lambda_c)\;{\rm exp}
\Biggl\lbrack\frac{\mu}{2}\Biggl{(}1-\sqrt{\frac{1-c}{25-c}}\;\Biggr{)}\Biggr\rbrack \; .
\end{equation}
For the minimal models of Eq.(\ref{eq:minseries}) the scaling variable is

\begin{equation}
\label{eq:minscalvar}
x=(\Lambda-\Lambda_c)\;{\rm exp}
\Biggl\lbrack\frac{m}{2m+1} \; \mu \Biggr\rbrack \; .
\end{equation}

The discrete formulation of $QG_2$ dates to 1982 and has proven to
be even more powerful than the continuum approach \cite{DisQG2}. It
is very rare in conventional field theories for the discrete formulation 
to be more flexible than the continuum treatment and 
the fact that it is so here certainly deserves notice.

In discrete $QG_2$ one replaces the 2d-manifold $\Sigma(\xi_1,\xi_2)$
by a set of $n$ discrete nodes $\{i\}$ and the metric $g_{ab}(\xi_1,\xi_2)$
by the adjacency matrix $C_{ij}$, where $C_{ij}=1$ if $i$ is connected to $j$
and $C_{ij}=0$ otherwise.
The connectivity $q_i$ of node $i$ is
\begin{equation}
\label{eq:conn}
q_i = \sum_j C_{ij} \; .
\end{equation}
The scalar curvature $R_i$ is
\begin{equation}
\label{eq:discurv}
R_i = 2\pi\frac{6-q_i}{q_i} \; .
\end{equation}
The gravity partition function becomes
\begin{equation}
\label{eq:disz}
Z(\Lambda,\mu) = \sum^{\infty}_0 e^{-\mu h} \sum^{\infty}_{n=0}
e^{-\Lambda n} Z_{h,n} \; ,
\end{equation}
where $Z_{h,n}$ for pure gravity is the number of distinct triangulations 
(${\cal T}$) of $n$ nodes with genus $h$ and $Z_{h,n}$ for models with matter
is symbolically 
\begin{equation}
\label{eq:dismatt}
Z_{h,n} = \sum_{{\cal T}} \int {\cal D}\lbrack \Phi \rbrack e^{-S_{MG}}
\end{equation}
for a generic matter field $\Phi$ coupled to gravity with action $S_{MG}$.
The density of states $Z_{h,n}$ is best computed by dualizing the 
triangulation to a $\Phi^3$ graph
and counting the number of distinct such connected graphs with $n$ vertices
that can be drawn, without crossing, on a surface of genus $h$ or greater.
To incorporate the topology of the graph one must promote $\Phi$ to an   
$N \times N$ matrix \`{a} la 't Hooft. 
The full partition function $Z(\Lambda,\mu)$ is, indeed, simply related
to the free energy of the corresponding matrix model:
\begin{equation}
\label{eq:matrixmodel}
Z(\Lambda,\mu) \equiv \frac{1}{N^2} {\rm  log} \; \zeta(N,g) \; ,
\end{equation}
where
\begin{equation}
\label{eq:matrixdef}
\zeta(N,g) = \int d^{N^2} \Phi \;{\rm exp} \lbrack -N\;{\rm Tr}\; ( \frac{1}{2}
\Phi^2 - \frac{g}{3} \Phi^3)\rbrack 
\end{equation}
with the identification
\begin{equation}
\label{eq:ident}
\frac{1}{N^2}={\rm e}^{-\mu} \quad \mbox{and} \quad g = {\rm e}^{-\Lambda}\; .
\end{equation}

\section{The $c=1$ Barrier}

As we have seen in the Introduction there is a good understanding of
how scaling dimensions of operators are modified by the coupling
of gravity to conformal field theories characterized by a central charge $c$
less than one. In several cases the models are exactly solvable and the 
sum over topologies is even possible in the double scaling limit $N
\rightarrow \infty$ and $\Lambda \rightarrow \Lambda_c$ with the scaling
variable $x$ of Eq.(\ref{eq:scalvar}) fixed \cite{dosc90}. 
When the central charge $c$
exceeds one the string susceptibility exponent $\gamma_s$, according to 
KPZ \cite{KPZ88}, becomes complex. This unphysical prediction is indicative
of an instability in the model. What is the true character of the 
theory for $c>1$? The $c=1$ point is analogous to the upper
critical dimension $d=4$ in the theory of phase transitions and, in fact,
there are known logarithmic violations of scaling at c=1 \cite{log}.

\begin{figure}[htb]
\vspace{9pt}
\epsfxsize=2.8in
\epsfbox{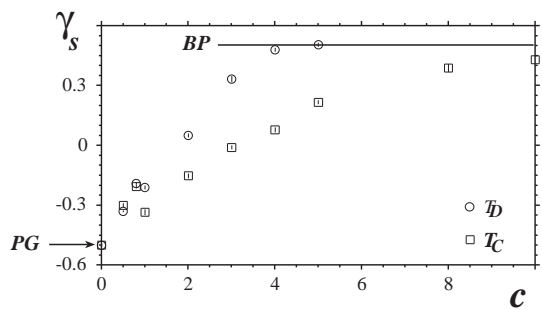}
\caption{$\gamma_s$ vs c for both degenerate ${\cal T_D}$ and combinatorial 
${\cal T_C}$ triangulations.}
\label{fig:gamc}
\end{figure}

From numerical simulations for $c>1$ the following picture has emerged.
Two basic classes of models have been carefully investigated.
The first consists of bosonic matter fields on dynamical
triangulations. In these models one finds a 
branched polymer {\bf BP} phase for $c$ large (typically $c\ge5$). 
The {\bf BP} phase is characterized by 
$\gamma_s=1/2$ and the proliferation of minimal neck baby universes
(mimbus). For smaller values of the central charge ($1<c<4$) the 
simulations indicate an intermediate phase with $0<\gamma_s<1/2$.
There is no apparent discontinuity in the vicinity of $c=1$.
The second class of models is multiple Ising models on
dynamical triangulations. The distinct species of spins couple
through the dynamical connectivity of the lattice.
For $n$ copies of Ising model there is a spin-ordering transition at a 
critical temperature. At the critical point one recovers a $c=n/2$ 
conformal field theory.
For $n>2$, but not too large, $\gamma_s$ increases smoothly from $0$ to $1/2$ 
with, again, no discontinuity near $n=2$ ($c=1$). For large $n$ the ordered and
disordered spin phases (both with pure gravity exponents $\gamma_s=-1/2$)
are separated by an intermediate {\bf BP} phase with $\gamma_s=1/2$ and
vanishing magnetization. The transition from the magnetized pure gravity 
phase to the disordered {\bf BP} phase is a branching transition with 
$\gamma_s=1/2$. This branching is one sign of the expected
instability discussed above. 
Similar phenomena are found for the q-state Potts model for q large.   
A plot of $\gamma_s$ as a function of the central charge $c$ from existing
simulations for two classes of triangulations (degenerate ${\cal T_D}$ and 
combinatorial ${\cal T_C}$) is shown in Fig.\ref{fig:gamc}.
 
Could there be a nontrivial infrared fixed point governing the
critical properties of the $c\ge1$ models? A renormalization group
scheme for matrix models, roughly analogous to the $\epsilon${-}expansion
about the upper critical dimension, was developed sometime ago by 
Br\'{e}zin and Zinn-Justin (BZ) \cite{BrZi92}. A typical example
is given by the $2^n$-matrix model formulation of the $n$-Ising spins
coupled to gravity. The BZ technique
consists of integrating out one row and one column of the matrix to
obtain an $N-1$ dimensional matrix. Changing $N$ is equivalent to changing 
the string coupling constant $e^{-\mu}$ of Eq.(\ref{eq:ident}). 
This induces a flow in the matrix model coupling 
which leads to definite renormalization group flows.
One can then look for fixed points.
The method is only qualitatively correct for the well-understood case
of $c<1$ minimal models but has the advantage it can be extended
to $c>1$. What does it tell us about $c>1$?

David's idea was to apply the BZ scheme to matrix models that include 
{\em branching} interactions. These interactions generate microscopic
wormholes that connect two macroscopic pieces of a Riemann surface. The
action for such models is given by
\begin{equation}
S_{N} (\Phi)  = N\;{\rm Tr} \left(\frac{\Phi^2}{2}-g\frac{\Phi^4}{4}\right)
-\frac{x}{2} {\rm Tr}^2\left(\frac{\Phi^2}{2}\right)
\end{equation}
The trace-squared (gluing) interaction (with coupling $x$) corresponds to
branching and is naturally generated at second order in perturbation
theory within a renormalization-group framework. Such models were first
treated by Das et al. \cite{Das90}. For pure gravity it was found that 
increasing the coupling $x$ leads to a new critical point $(g_c, x_c)$ in the 
$(g,x)${--}plane separating a large-$x$ {\bf BP} phase ($\gamma=1/2$) from 
the pure gravity phase ($\gamma=-1/2$). 
At the (branching) transition $\gamma_s=1/3$.
This result extends to the case of the $m-th$ minimal model. In this case
the branching critical point has $\gamma_s=1/(m+1)$. Applying the
BZ RG method to this case David found the RG flows
\begin{equation}
\begin{array}{l}
\beta_{g} = g-6g{^2} - 2g x \\
\beta_{x} = 2x - 3x{^2} - 6g x
\end{array}
\label{eq:rgflow}
\end{equation}

For $c>1$ the only true fixed point is the {\bf BP} fixed point. 
But the influence of 
the $c<1$ fixed points is still felt for $c$ in the neighbourhood of $1$.
For $c<1$ the scaling dimension of the field $x$ (corresponding to the 
renormalized coupling $x$) is positive and wormholes are irrelevant. 
At $c=1$ the branching interaction is marginal and the matter-gravity fixed 
point merges with
the branching fixed point. For $c>1$ wormholes are relevant and the 
matter-gravity and branching fixed points become complex conjugate pairs 
off the real
axis. As a result there are exponentially enhanced crossover effects which
imply that an exponential fine-tuning of couplings
\begin{figure}[htb]
\vspace{9pt}
\epsfxsize=2.8in
\epsfbox{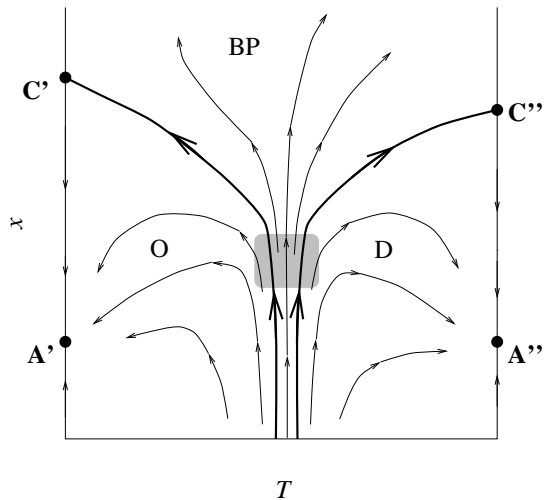}
\caption{Schematic RG flows for $n>2$ ($c>1$) multiple-Ising models from 
\protect\cite{Davi97}.}
\label{fig:rgflow}
\end{figure}

\begin{equation}
\label{eq:finetune} 
|g-g{_c}| \sim {\rm exp} \left(-\frac{1}{\sqrt{c-1}}\right)
\end{equation} 
is necessary to see the flow to the true {\bf BP} fixed point. 
Without this fine-tuning flows appear similar to those for $c<1$.  
Fig.~\ref{fig:rgflow} shows a schematic RG flow for the generalized 
$n>2$ ($c>1$) multiple-Ising model with branching interactions.
In the shaded region flows are similar to the 
case $c<1$ unless the temperature is fine-tuned. {\bf A'} and {\bf A''}
denote pure gravity fixed points and {\bf C'} and {\bf C''} denote
branching critical points. The regions $O$ and $D$ are ordered and disordered 
spin phases respectively. The line $x=0$ corresponds to the original
discretized model of $n$-Ising spins on a dynamically triangulated lattice. 
On this line there is no spin-ordering transition {--} only the branching
transition to the {\bf BP} phase.   

\begin{figure}[htb]
\vspace{9pt}
\epsfxsize=2.8in
\epsfbox{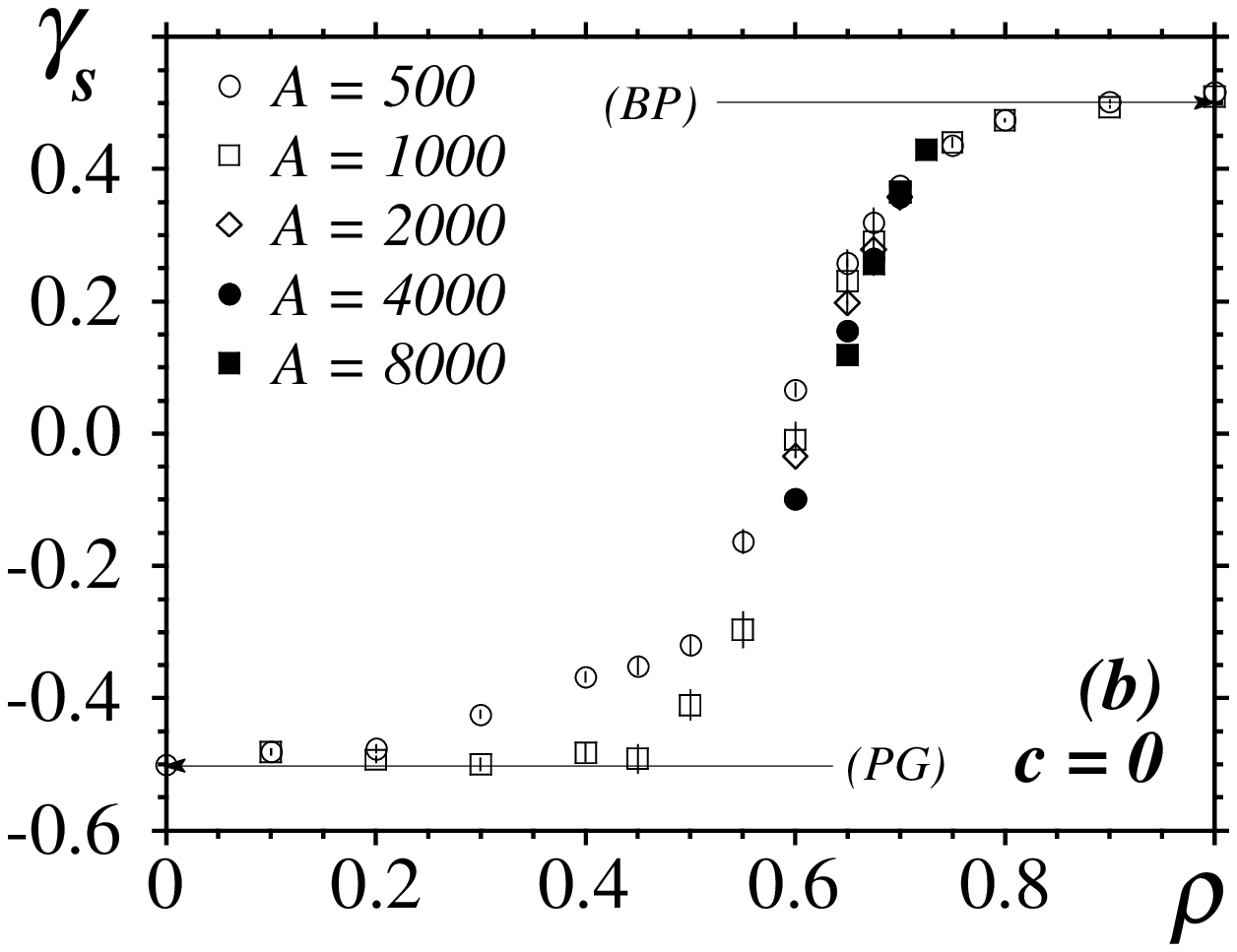}
\caption{$\gamma_s$ vs $\rho$ for $c=0$ from \protect\cite{ThPe97}.}
\label{fig:gamc0}
\end{figure}

\begin{figure}[htb]
\vspace{9pt}
\epsfxsize=2.8in
\epsfbox{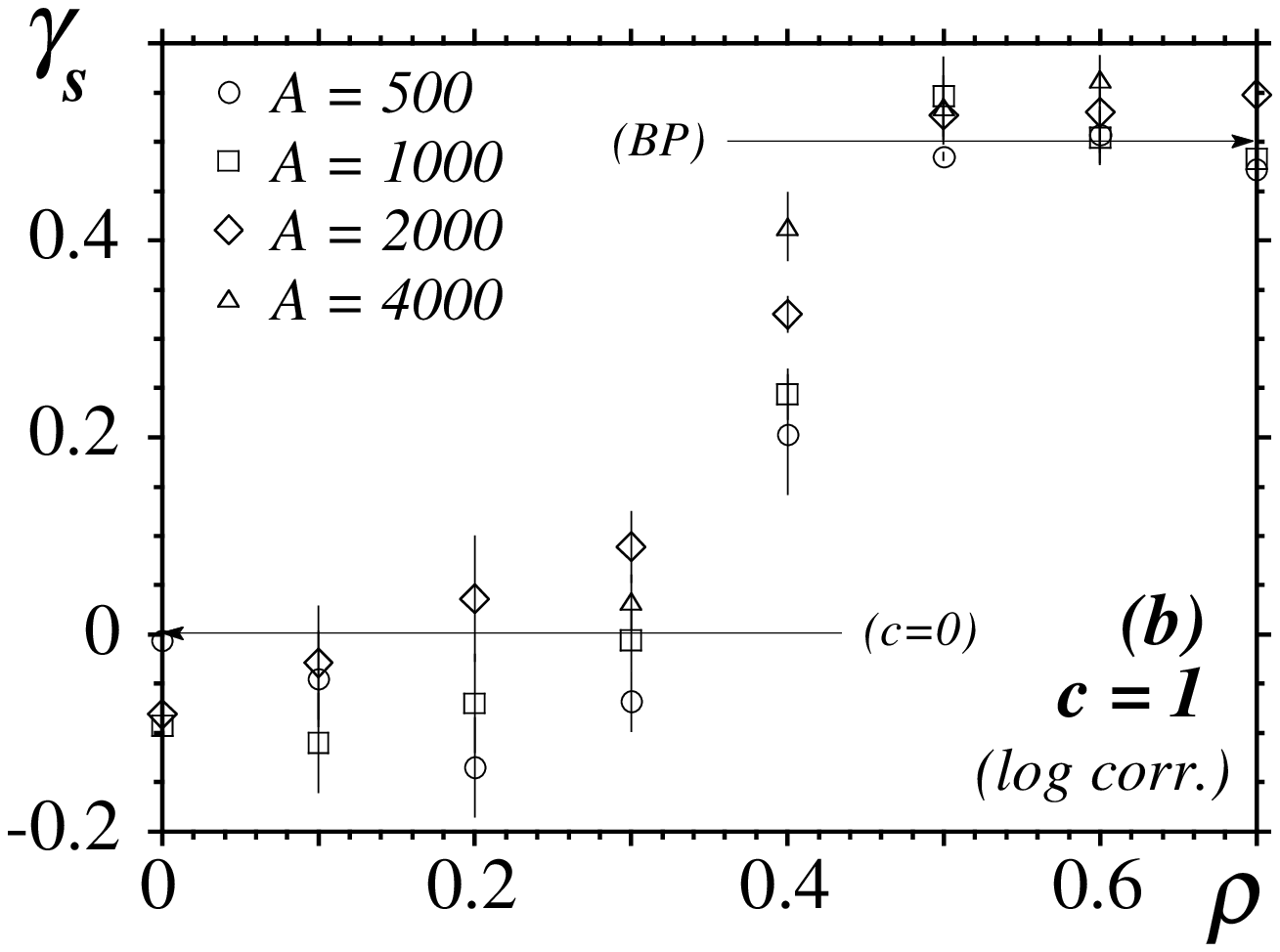}
\caption{$\gamma_s$ vs $\rho$  for $c=1$ from  \protect\cite{ThPe97}.}
\label{fig:gamc1}
\end{figure}

\begin{figure}[htb]
\vspace{9pt}
\epsfxsize=2.8in
\epsfbox{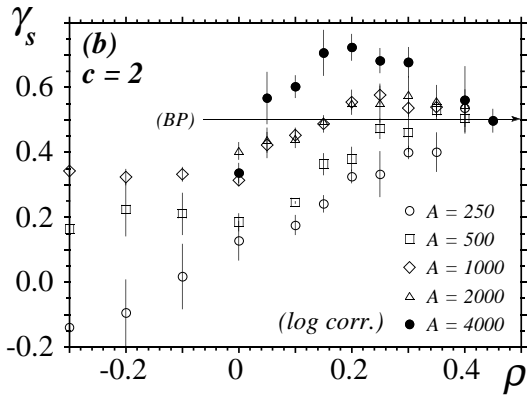}
\caption{$\gamma_s$ vs $\rho$ for $c=2$ from \protect\cite{ThPe97}.}
\label{fig:gamc2}
\end{figure}

This nice picture is consistent with the existing numerical results and 
may also be tested numerically.
A first step was reported at this meeting \cite{ThPe97}. These authors
study the partition function
\begin{equation}
Z_A=\sum_{T\epsilon {\cal T}} {\rm e}^{\rho n_m}Z_m
\end{equation}
where $\rho$ is a chemical potential, $n_m$ is the number of minimal necks 
and $Z_M$ is a standard matter action for multiple Gaussian fields. The cases 
studied are zero ($c=0$), one ($c=1$) and two ($c=2$) scalar fields. They look
for a transition to the {\bf BP} phase at a finite critical chemical potential
$\rho_c$. The clearest results come from an analysis of $\gamma_s$ and are 
shown in Figs.~\ref{fig:gamc0}{--}\ref{fig:gamc2}. For
$c=0$ and $c=1$ $\gamma_s$ changes sharply from $\gamma\sim -1/2$ to
$\gamma\sim 1/2$ at a definite $\rho_c$ with $\rho_c \sim 0.6$ ($c=0$)
and $\rho_c \sim 0.4$ ($c=1$). For $c=2$ one finds instead a smooth 
volume-dependent cross-over to the {\bf BP} phase. 
Furthermore the results are consistent with the
model being only in the {\bf BP} phase in the infinite volume limit. These 
interpretations are supported by an analysis of the specific heat. For $c=0$ 
and $c=1$ one sees a definite phase transition but for $c=2$ no
self-consistent critical exponents can be extracted from finite-size
scaling of the specific heat curves.

\section{Fractal Dimension} 

\begin{table*}[hbt]
\setlength{\tabcolsep}{1.5pc}
\newlength{\digitwidth} \settowidth{\digitwidth}{\rm 0}
\catcode`?=\active \def?{\kern\digitwidth}
\caption{The fractal dimension for $c\le1$ models: theory and simulations.}
\label{table:dh}
\begin{tabular*}{\textwidth}{@{}l@{\extracolsep{\fill}}rrrrrr}
\hline
\multicolumn{6}{c}{$d_h$}\\
\hline
\multicolumn{1}{l}{$c=-2$}        &
\multicolumn{1}{r}{$c=0$}        & 
\multicolumn{1}{r}{$c=1/2$}        & 
\multicolumn{1}{r}{$c=4/5$}        &
\multicolumn{1}{r}{$c=1$}        & 
\multicolumn{1}{c}{Method} \\
\hline
2         &  4           &  6           &  10          & $\infty$ &
Theory:$\;$Eq.(\ref{eq:dhformone}) \\
3.562     &  4           &  4.21        &  4.42        & 4.83     & Theory:$\;$Eq.(\ref{eq:dhformtwo}) \\
3.58(4)  & 3.58{--}4.20     & 3.95{--}4.35     & 4.00{--}4.55      & 3.8{--}4.4 
& Simulations\\
\hline
\end{tabular*}
\end{table*}

Our current understanding of the purely spacetime aspects of $QG_2$ coupled
to matter is much less complete than our knowledge of the effects of gravity
on the matter fields and critical behaviour. Of basic interest is the 
intrinsic Hausdorff dimension of the typical surface appearing in the 
ensemble of random surfaces. The Hausdorff dimension is defined by
\begin{equation}
d_H=\frac{\log A}{\log r} \; ,
\end{equation}
where $A$ is the area of the surface and $r$ is some appropriate measure 
of the geodesic size of the surface. There is a considerable variety of 
alternative definitions of $d_H$. In fact $A$ and $r$ here may be replaced by
any reparameterization-invariant observables with dimensions of area and 
distance respectively.

For the case of pure $QG_2$ $d_H$ is known to be 4 \cite{KKMW93,AmWa95}. 
This result employs a transfer matrix formalism to study the evolution of
loops on the surface. One may also use a string field theory for non-critical 
strings \cite{IsKa93} to calculate $d_H$. Perhaps the simplest approach is to 
formulate the theory on a disk with boundary length $\ell$ and to use $\ell$ as
a ruler for determining the scaling of both $A$ and $r$. In the string field
theory approach an ADM-type gauge is chosen in which geodesic distance plays
the role of time. Together with a choice of the string field theory
Hamiltonian this determines the scaling of geodesic distance $r$ with
boundary length $\ell$ to be
\begin{equation}
r \sim \ell^{1/m} \; ,
\end {equation}
for the $m-th$ minimal model coupled to $QG_2$. The scaling of $A$ vs. $\ell$ 
may be determined by standard matrix model calculations. If we assume that the
area $A$ scales canonically as $\ell^2$ we conclude that
\begin{equation}
A \sim r^{2m},
\end {equation}
implying $d_H = 2m$. This result may be equivalently written as
\begin{equation}
\label{eq:dhformone}
d_H = \frac{2}{|\gamma_s|} = \frac{24}{1-c+{\sqrt{(25-c)(1-c)}}}\; .
\end{equation}
A completely different result is obtained by studying the diffusion of a 
fermion with the methods of Liouville theory \cite{KSW93}. This gives
\begin{equation}
\label{eq:dhformtwo}
d_H = -2\frac{\alpha_1}{\alpha_{-1}}=2 \times 
\frac{\sqrt{25-c}+\sqrt{49-c}}{\sqrt{25-c}+{\sqrt{1-c}}} \; ,
\end {equation}
where $\alpha_n$ is the gravitational dressing of a $(n+1,n+1)$ primary
spinless conformal field.

In Table~\ref{table:dh} we list the predictions from these two formulae 
together with the results from numerical simulations \cite{Nudh95,Anag97}.

\begin{figure}[htb]
\vspace{9pt}
\epsfxsize=2.8in
\epsfbox{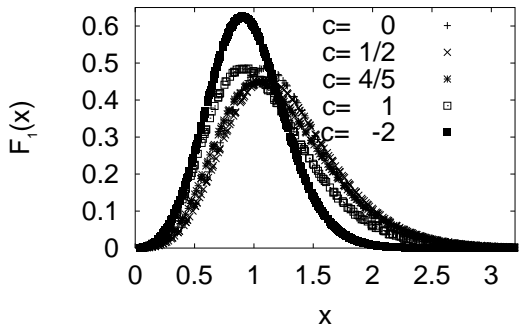}
\caption{Scaling fits for the two point function for $QG_2$ from 
\protect\cite{Anag97}.}
\label{fig:dhscal}
\end{figure}

Both the analytic predictions discussed above, as well as the exact solution,
agree on $d_H=4$ for pure $QG_2$. A detailed numerical investigation of
the fractal dimension, determining both the scaling of two-point functions 
defined in terms of geodesic distance and the behavior of the loop length 
distribution function, was presented in this meeting \cite{Anag97}. 
For $0<c\leq{1}$ $d_H$ is found to be very close to 4, in agreement with 
earlier simulations \cite{Nudh95}. 
An example of the excellent scaling curves obtained 
is shown in Fig.~\ref{fig:dhscal}.
The current best numerical results therefore do not agree with either of the 
analytic predictions discussed above. There is not yet complete
agreement, to high precision, between the various methods of determining
$d_H$ and so one cannot confidently rule out both theories. Further work
is highly desirable to consolidate the result that $d_H$ for $0<c<1$
is independent of the matter. 
The subtlety of the problem when Ising matter is included may be
seen in \cite{BJT97}. The scaling of area versus boundary length depends on
the precise order in which the infinite-volume limit is taken with respect
to the tuning to the Ising critical temperature. It is possible to obtain
$d_H=4$ if the infinite-volume thermodynamic limit is taken with $T\neq{T}_{c}$
followed by tuning to the spin-ordering transition. Finally large-scale
numerical results for $c=-2$ \cite{AAIJKWY97}, made possible by recursive
sampling of the space of graphs for this topological model, are in
excellent agreement with the Liouville prediction $d_H=3.56$.

\section{Higher Dimensional Simplicial Gravity}

Since the DT approach to quantum gravity is very successful in two 
dimensions it is natural to explore its implications in
higher dimensions. There has, in fact, been considerable effort
in this direction in recent years.
Consider, to  begin, the case of pure Einstein-Hilbert gravity
in $D$ dimensions for $D=3$ or $4$. 
The functional integral to be evaluated is
\begin{equation}
\label{eq:part}
Z = \sum_{{\cal M} \in {\rm Top}} \int_M \frac{D \lbrack g \rbrack} {{\rm Vol}({\rm Diff})}
\,{\rm e}^{-S(g)},
\end{equation}
where $M$ is the spacetime manifold with topology chosen from the class 
{\em Top} 
and the action $S(g)$ is given by
\begin{equation}
\label{eq:ehaction} 
S(g) = \int d^4x\;\sqrt{g}\;(\Lambda - \frac{R}{16\pi G}).
\end{equation}

In the following we will restrict ourselves to fixed topology since
there is little rigorous understanding of the meaning of the
sum over of topologies in the general higher dimensional case. 
The continuum integral over diffeomorphism-inequivalent
metrics is replaced in the DT approach by a discrete sum over all
possible cellular decompositions (gluings) of
$D${-}simplices along their $D-1${-}faces with the simplicial manifold
requirement that the neighbourhood of each vertex is a $D$-ball and the
DT constraint that all edge lengths are fixed. So far the majority
of numerical results have been limited to the fixed topology
of the sphere. To be specific let us concentrate on the $4$-dimensional 
case of the four-sphere $S^4$.

\begin{figure}[htb]
\centerline{\psfig{file=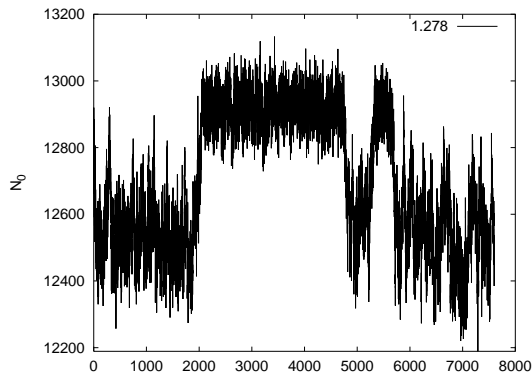,width=2.8in,angle=270}}
\caption{4d Monte Carlo time history of $N_0$ for $N_4=64,000$ and
$k_2=1.278$ from \protect\cite{BdeB96}. The horizontal units are 100 sweeps.}
\label{fig:firstorder}
\end{figure}

\begin{figure}[htb]
\vspace{9pt}
\epsfxsize=2.5in
\epsfbox{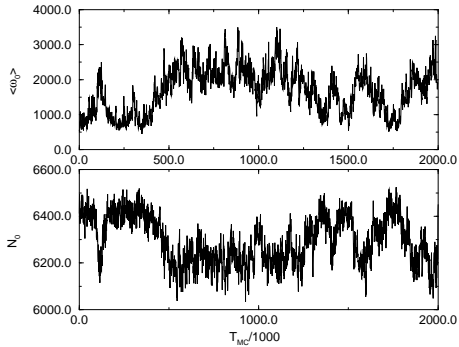}
\caption{Correlation of vertex number $N_0$ with singular vertex volume from
\protect\cite{CRK97}.}
\label{fig:singvertex}
\end{figure}

The free global variables at our disposal are the numbers $N_i({\cal T})$ of 
$i$-dimensional simplices in a given triangulation ${\cal T}$ ($i=0,1,..,4$).
These 5 parameters are constrained by two Dehn-Sommerville relations
\begin{equation}
\label{eq:dehn}
\sum^4_{i=2k-1} (-1)^i \pmatrix{i+1\cr 2k-1} N_i({\cal T}) = 0 \quad k=1,2 
\end{equation}
together with the Euler relation 
\begin{equation}
\label{eq:euler}
\chi\left(S^4\right) = \sum_0^4 (-1)^i \; N_i({\cal T}) = 2 \; . 
\end{equation}
These three relations leave two independent variables which we may take to be
$N_2$ and $N_4$. These are the discrete analogues of the mean scalar
curvature and the volume. The Einstein-Hilbert action Eq.(\ref{eq:ehaction})
then becomes
\begin{equation}
\label{eq:discreteeh}
Z\lbrack k_2,k_4 \rbrack = \sum_{{\cal T}(S^4)} {\rm exp} \lbrack -k_4
N_4({\cal T}) + k_2 N_2({\cal T}) \rbrack 
\end{equation}
where $k_4$ is the discrete cosmological constant and  $k_2$ is the
discrete inverse Newton's constant.
In practice (almost) fixed $N_4$ (volume) systems are usually simulated 
by adding a constraint that restricts the volume to be near a target volume.
One is then really approximating  the fixed volume partition function
\begin{equation}
\label{eq:fixedvol}
Z\lbrack k_2,N_4 \rbrack = \sum_{{\cal T}(S^4)} {\rm exp} \lbrack k_2 
N_2({\cal T}) \rbrack \; .
\end{equation}
From extensive numerical simulations it has emerged that the
system described by the partition function of Eq.(\ref{eq:fixedvol})
has two distinct phases. For $k_2$ small (strong coupling) the system
is crumpled (infinite Hausdorff dimension) and the mean scalar curvature
$\langle R \rangle$ is negative. For $k_2$ large (weak coupling)
the system is elongated (branched-polymer like) with Hausdorff dimension 2
and positive mean scalar curvature of order the volume. In the crumpled
phase generic triangulations contain one singular one-simplex with
two singular vertices \cite{HIN95,CTKR96}. 
These singular simplices are connected to an extensive
fraction of the volume of the simplicial manifold. The local volume associated
with the singular one-simplex grows like $V^{2/3}$ while the local volume 
associated with the singular vertices grows like $V$ itself. 
The appearance of singular structures is generic to simplicial DT
gravity in dimensions $D\ge4$.
For $D\ge4$ one finds a singular $D-3$ simplex (with 
local volume of order $V^{2/3}$) and singular sub-simplices 
(with local volume of order $V$).

It is now clearly established that there is a {\em first order}
phase transition connecting the crumpled phase with the branched polymer phase
in both 3 \cite{QG3} and 4 dimensions \cite{BBKP96,BdeB96}.
This is seen dramatically in the time series of Monte Carlo sweeps shown in 
Fig.~\ref{fig:firstorder} \cite{BdeB96}. 
A finite-size scaling analysis of the variance of the $N_2$ 
(or equivalently $N_0$) fluctuations also reveals a maximum which
grows linearly with the system volume {---} a classic signal
of a first-order transition.   
It requires both large volumes (order $64,000$) and long simulations (order  
$10^6$ sweeps) to see the first order nature of the transition emerge clearly
in four dimensions.
It is also becoming clear that the transition itself is closely
connected to the formation of the singular simplices \cite{CRK97}.
This is demonstrated in Fig.~\ref{fig:singvertex}.

Many characteristics of the elongated phase are analogous to those of 
branched-polymers \cite{AmJu95} and may be reproduced by simple and elegant
statistical mechanical models of branching \cite{BrPo}.
With a suitable choice of ensemble this class of models also
possesses a discontinuous transition \cite{BBJ97}.  
Thus the weak coupling phase of the theory may reflect more about the 
combinatorial nature of the simplicial DT action than about the nature of 
gravity itself. This remains to be seen.

The lack of a continuous transition for DT simplicial gravity in   
higher dimensions has been a deterrent for recent
progress in the field. From a traditional point of view
this absence of a critical point implies that the model has no continuum
limit we could call continuum quantum gravity.
At least three viewpoints are possible at this stage.
It may be that the theory is fundamentally discrete and that
{\em no} continuum limit should be taken. This viewpoint is advocated
in different ways in several other approaches to quantum gravity such as the
causal sets formulation of Sorkin \cite{Sorkin} and the theory of spin 
networks \cite{Smolin}.
Alternatively the theory might only have an interpretation as an effective
theory valid over a limited range of length scales.
Finally it may be that pure DT simplicial gravity is ill-defined but that 
models with appropriate matter or modified measures \cite{BrMa93}
possess critical points and admit a suitable continuum limit.
This latter approach was discussed in this meeting by Izubuchi 
for $QG_3$ \cite{Izu97}. One modifies the action by adding terms
that effectively enhance higher curvature fluctuations {---} these
correspond to changing the measure locally by powers of $\sqrt{g}$. 
There is some evidence that by tuning this extra term one can
soften the transition to a continuous one at finite volumes.
It is very likely though that the effect disappears in the infinite-volume
limit. This issue needs further clarification.

\section{Membranes}

The theory of random surfaces with the addition of an extrinsic curvature
to the action has a direct connection with the statistical
mechanics of flexible membranes \cite{MemRev}. 
Physical membranes are two-dimensional
surfaces fluctuating in three embedding dimensions. 
The simplest examples of 2-dimensional surfaces are strictly planar systems
called films or monolayers. These alone are surprisingly complex systems.
But when they also have the freedom to fluctuate, or bend,
in the embedding space it is a considerable challenge to determine 
their physical properties. Two broad classes of membranes have been 
extensively studied. Membranes with fixed connectivity (bonds are never
broken) are known as crystalline or tethered membranes. Membranes
with dynamical connectivity (relatively weak bonds which are free to break 
and rejoin) are known as fluid membranes.  

The simplest action for self-intersecting (non-self-avoiding)
fluid membranes resembles that 
of the Polyakov string with extrinsic curvature (bending energy). 
Since the beta function for the inverse-bending rigidity 
(where the bending rigidity is the coupling to the extrinsic curvature)
is well-known to be asymptotically free at one loop, the bending
rigidity is irrelevant in the infrared and self-intersecting
fluid membranes are commonly believed to exist only in a crumpled phase.

Crystalline membranes, on the other hand, have a non-linear 
coupling between elastic (phonon) interactions and bending 
fluctuations which drives a phase transition from a high-temperature 
crumpled phase to a low-temperature orientationally ordered (flat) phase.

\begin{figure}[htb]
\vspace{9pt}
\centerline{\psfig{file=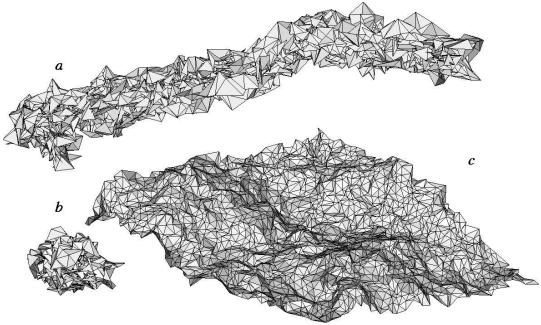,width=2.8in,clip=}}
\caption{The three phases of anisotropic membranes:(a) tubular (b) crumpled 
and (c) flat.}
\label{fig:tubules}
\end{figure}

Much can be learned about membranes by applying the techniques and experience
of lattice gravity simulations to these condensed matter/biological
problems. This is illustrated by some recent \cite{BFT97}
large-scale Monte Carlo simulations of anisotropic crystalline membranes 
{---} these are systems in which the
bending or elastic energies are different in different directions.
We were able to verify for the first time the predicted existence
\cite{RT95} of a novel {\em tubular} phase in this class of membranes.
This phase is intermediate between the flat and crumpled phases {---}
it is extended (flat) in one direction but crumpled in the transverse
direction. Correspondingly there are two phase transitions: the
crumpled-to-tubular transition and the tubular-to-flat transition.
Thermalized configurations from the three phases of anisotropic crystalline 
membranes are shown in Fig.~\ref{fig:tubules}. 
These simulations employed a variety of improved Monte Carlo algorithms
such as hybrid overrelaxation and unigrid methods \cite{ThFa97}.
An even more challenging problem is to incorporate the self-avoidance found 
in realistic membranes \cite{BG97,RT297}.

Finally the full physics of fluid membranes, in which dynamical connectivity
may be modeled by dynamical triangulations, is a problem of limitless 
challenges which ties together common techniques in lattice gravity and the 
burgeoning field of soft condensed matter physics \cite{Lub96}.

\section{Future Challenges}

There are many challenges facing the program of simplicial lattice gravity.
I will give here a few outstanding problems and possible future
directions.

\vspace{0.4cm}

\noindent $\bullet$ {\em Topology Change}

\vspace{0.4cm}

Most of the numerical simulations in the field 
have been on spacetimes with fixed topology. From the functional integral
point of view it is more natural (and perhaps essential) to allow the
topology of spacetime to fluctuate. In $2d$ it is even possible 
to perform analytically the sum over all topologies in the double scaling 
limit. A preliminary investigation of a $4d$ dynamical triangulation model
of Euclidean quantum gravity with fluctuating topology was made some
time ago by de Bakker \cite{BdeB95}. 

\vspace{0.4cm} 

\noindent $\bullet$ {\em Supersymmetry}

\vspace{0.3cm}

Whether or not it has physical relevance it is clear that
one of the most powerful ideas in particle physics/quantum field
theory/string theory at present is supersymmetry. Supersymmetry
severely constrains the analytic structure of any model.
Almost no progress has been made in formulating or simulating 
supersymmetric simplicial gravity or supersymmetric random
surfaces. If we are ever to make contact with critical or non-critical
superstrings and recent developments like duality relations 
this will be essential.

\vspace{0.3cm} 

\noindent $\bullet$ {\em Classical Limit}

\vspace{0.3cm}

Although simplicial quantum gravity does provide a non-perturbative
definition of quantum gravity in four dimensions there is no understanding
of how classical gravity emerges in the long-wavelength limit and indeed 
of how perturbative scattering amplitudes are reproduced in this framework.

\vspace{0.3cm} 

\noindent $\bullet$ {\em Triangulation Class}

\vspace{0.3cm}

At present there is no classification of the precise class of graphs
that result in KPZ, rather than Onsager, critical exponents
when matter is coupled to a dynamical lattice ($QG_2$). 
For example it has been shown \cite{BCT97} that MDT 
({\em minimal dynamical triangulation}) models in which the local
coordination number is restricted to be 5, 6 or 7 still result
in KPZ exponents.  

\vspace{0.3cm} 

\noindent $\bullet$ {\em Fractal Dimension}

\vspace{0.3cm}

Is the fractal dimension of all $0<c<1$ matter coupled to $QG_2$ really four?

\vspace{0.3cm} 

\noindent $\bullet$ {\em Interpretation of $QG_4$}

\vspace{0.3cm}

What is the correct interpretation of $4d$ simplicial quantum gravity
given the first order transition from the strong to weak coupling phases?
What is the proper mathematical framework for $QG_4$ and how much analytic 
progress can be made? Recent progress in this direction is reviewed in 
\cite{ACM97}.

\vspace{0.3cm}

\noindent $\bullet$ {\em Renormalization Group}

\vspace{0.3cm}

Recently renormalization group (RG) methods have been developed for
pure simplicial gravity and for simple matter coupled to gravity
\cite{RG}. There is no rigorous understanding of the principles
behind the success of these methods and further
improvement of the technique is highly desirable.
Further progress in this direction to enable, for example,
the direct computation of the beta function for random surfaces with
extrinsic curvature would be very nice.

I would like to acknowledge Kostas Anagnostopoulos, Simon Catterall, 
Marco Falcioni and Gudmar Thorleifsson for extensive discussion 
on many issues treated in this talk.

\end{document}